\documentclass[aps,prb,twocolumn,floats]{revtex4}

\usepackage{ifthen}
\usepackage{ifpdf}

\usepackage{latexsym}
\usepackage{amsmath}
\usepackage{amssymb}
\usepackage{bm}
\usepackage{wasysym}

\ifpdf
\usepackage{graphicx}
\usepackage{epstopdf}
\else
\usepackage{graphicx}
\usepackage{epsfig}
\fi

\newcommand{\tbox}[1]{\mbox{\tiny #1}}

\newcommand{\be}[1]{\begin{eqnarray}\ifthenelse{#1=-1}{\nonumber}{\ifthenelse{#1=0}{}{\label{e#1}}}}
\newcommand{\ee}{\end{eqnarray}}

\newcommand{\hide}[1]{}

\newcommand{\sect}[2]{\section{#2}}

\begin{document}

\title{From the Kubo formula to variable range hopping}

\author{Doron Cohen}

\affiliation{
Department of Physics, Ben-Gurion University, Beer-Sheva 84105, Israel
}

\begin{abstract}
Consider a multichannel closed ring with disorder.
In the semiclassical treatment its conductance is
given by the Drude formula. Quantum mechanics
challenge this result both in the limit of strong disorder
(eigenstates are not quantum-ergodic in real space)
and in the limit of weak disorder
(eigenstates are not quantum-ergodic in momentum space).
Consequently the analysis of conductance 
requires going beyond linear response theory,  
leading to a resistor network picture of 
transitions between energy levels. We demonstrate 
that our semi-linear response theory provides a firm 
unified framework from which the ``hopping" phenomenology 
of Mott can be derived. 
\end{abstract}


\maketitle


\sect{(0)}{Introduction}

The theory for the conductance of {\em closed} mesoscopic rings 
has attracted a lot of interest 
\cite{rings,debye1,debye2,debye3,IS,IS1,loc1,loc2,loc3,G1,G2,kamenev}.  
In a typical experiment \cite{orsay} 
a collection of mesoscopic rings are driven by 
a time dependent magnetic flux $\Phi(t)$ which creates 
an electro-motive-force (EMF) ${-\dot{\Phi}}$ in each
ring. Assuming that Ohm's law applies, the induced current 
is ${I=-G\dot{\Phi} }$ and consequently Joule's law gives 
\be{1}
\dot{\mathcal{W}} 
\ \ = \ \ 
G\,\dot{\Phi}^2 
\ \ = \ \ 
\mbox{rate of energy absorption} 
\ee
where $G$ in this context is called the conductance.
For diffusive rings the Kubo formula 
leads to the Drude formula for $G$.  
A major challenge in past studies 
was to calculate the weak localization 
corrections to the Drude result, 
taking into account the level statistics  
and the type of occupation \cite{G1,G2,kamenev}. 
It should be clear that these corrections 
do not challenge the leading order Kubo-Drude result.


One wonders what happens 
to the Drude result if the disorder becomes 
weak (ballistic case) or strong 
(Anderson localization case). 
In both cases the individual eigenfunction 
become non ergodic: a typical eigenfunction 
do not fill the whole accessible phase space. 
In the ballistic case a typical eigenfunction  
is not ergodic over the open modes in momentum space,  
while in the strong localization case 
it is not ergodic over the ring in real space.


Lack of quantum ergodicity implies that 
the perturbation matrix 
is very structured and/or sparse. 
Consequently the calculation of~$G$ 
requires a non-trivial extension  
of linear response theory (LRT). 
Such extension has been proposed in Ref.\cite{kbr}  
and later termed ``semi linear response theory" (SLRT) \cite{slr}.    
Within SLRT it is still assumed that the   
transitions between levels are given 
by Fermi-golden-rule, but a resistor network 
analogy is used in order to calculate 
the overall absorption. 
In order to have non-zero absorption we must 
have {\em connected sequences of transitions}.  
Thus the calculation of the energy absorption 
in Eq.(\ref{e1}) is somewhat similar to solving 
a percolation problem. The ``percolation" is in 
energy space rather than in real space.


In a previous work we have worked out results for $G$ 
in the case of a ballistic ring \cite{bls}. In this work we 
would like to explore the other extreme case of strong disorder. 
If we go to the literature we find a strange twist. 
For the AC conductance Mott \cite{mott1} (see also \cite{sivan}) 
has used the Kubo formula in order to predict 
$\sim \omega^2 |\log(\omega)|^{d{+}1}$ dependence of 
the conductivity, where $d=1,2,3$ is the dimensionality 
of the sample. On the other hand for the DC calculation 
Mott has abandoned the Kubo formalism and has adopted 
a phenomenological variable range hopping (VRH) picture \cite{mott2}, 
which can be regarded as an approximation 
for a more elaborated (but still phenomenological) 
resister network picture \cite{miller,ambeg,pollak}.  
The ``ad hoc" approach to hopping is obviously bothering. 
For example it is not clear  whether the effect 
of low frequency noisy driving should treated 
like ``low temperature" or like "small frequency" \cite{pc}. 
The main purpose of this Letter is to explain 
that the hopping picture is a natural outcome of SLRT, 
and hence can be regarded as a natural extension of LRT.    
This automatically resolves such conceptual problems, 
and opens the way for further refinements of the theory.


The outline of this Letter is as follows:
We summarize the LRT and the SLRT recipes 
for the calculation of the conductance. 
Then we demonstrate that all the known results 
for the conductance can be derived from the 
same theoretical framework, 
without any extra assumptions.

\sect{(1)}{The model parameters:}
We consider a ring of length $L$ and mean free path $\ell$.
The particles are non-interacting ``spinless" electrons 
with charge~$e$. The Fermi energy and the Fermi 
velocity are $E_F$ and $v_{\tbox{F}}$ respectively. 
The one-particle density of states at the Fermi energy is 
\be{2}
\varrho_{\tbox{F}} = \mbox{GeometricFactor} \times \mathcal{M} \frac{L}{\pi\hbar v_{\tbox{F}}}
\ee
The number of open modes $\mathcal{M}$ reflects the cross 
section of the ring. The GeometricFactor depends on the 
dimensionality $d=1,2,3$ and equals unity for $d=1$ network systems. 
In what follows, for sake of presentation, and without loss of generality, 
we set in all expressions the $d=1$ geometric factor, 
and use units such that $\hbar=1$.

The mean level spacing is $\Delta=\varrho_{\tbox{F}}^{-1}$.
This is the "small" energy scale. 
We also have the ``Thouless energy" or its equivalent 
which we denote as $\Delta_b$. 
For ballistic ring $\Delta_b = \mathcal{M}\Delta $ 
is associated with the time scale $L/v_{\tbox{F}}$,  
while in the diffusive regime ${\Delta_b=(\ell/L)\mathcal{M}\Delta}$ 
is associated with the ergodic time.
In the strong disorder regime we shall see 
that the relevant time scale for our analysis 
is the breaktime $t^*$ which is related to the 
localization length~$\ell_{\xi}$.

Within the framework of both  LRT and SLRT 
it is essential to realize that the combined 
effect of the driving and the environment 
is to {\em ``broaden"} the energy levels 
and to induce {\em ``relaxation"}. Accordingly we 
distinguish between~$\Gamma$  which can be 
interpreted as the dephasing or decoherence rate  
(analogous to $1/T_2$ in NMR studies), 
and $\gamma_{\tbox{rlx}}$ which is the relaxation rate 
(analogous to $1/T_1$ in NMR studies).
We assume that $\Gamma$ is much larger than $\Delta$  
but much smaller compared with $\Delta_b$. 
We also assume that the driving source  
is "essentially DC driving". This means that  
the driving frequency is much smaller compared with $\Delta_b$.
LRT applies whenever the driven transitions 
are much slower compared with the relaxation rate. 
SLRT is an extension of LRT that applies 
in the opposite circumstances, 
namely if the relaxation process can be neglected 
during the time that the absorption rate 
get stabilized. This is explained in length in\cite{kbr},
and will be emphasized again in a later paragraph.

\sect{(2)}{The semiclassical LRT calculation:}
The semiclassical Kubo formula once expressed 
with the velocity-velocity correlation function is
\be{3}
G = \varrho_{\tbox{F}} \times \frac{1}{2} 
\left(\frac{e}{L}\right)^2
\int_{-\infty}^{\infty} \langle v(t) v(0) \rangle dt
\ee
where $v$ is the velocity of the particle along the ring.
For the purpose of later reference we note that Eq.(\ref{e1})
[with Eq.(\ref{e3}) substituted] is easily generalized 
to the case where the driving is noisy. 
Instead of a multiplication by $\dot{\Phi}^2$ we have in general 
an integration over the all the Fourier components of the driving. 
Namely, 
\be{4}
\dot{\mathcal{W}} 
= \varrho_{\tbox{F}} \times \frac{1}{2} 
\left(\frac{e}{L}\right)^2
\int_{-\infty}^{\infty}  
d\omega \,
\tilde{F}(\omega) \, 
\tilde{C}(\omega)
\ee
where $\tilde{C}(\omega)$ is the Fourier transform 
of the velocity-velocity correlation function, 
and $\tilde{F}(\omega)$ is the power spectrum of $\dot{\Phi}(t)$.

The Kubo formula parallels the expression for 
the spatial diffusion coefficient, which is the integral 
over the velocity-velocity correlation function: 
\be{5}
\mathcal{D}
= \frac{1}{2} 
\int_{-\infty}^{\infty}   
\langle v(t) v(0) \rangle dt
\ee
In the case of pure DC driving (${F(\omega)=\dot{\Phi}^2\delta(\omega)}$)
one obtains the Einstein relation:
\be{6}
G=\varrho_{\tbox{F}}\left(\frac{e}{L}\right)^2\mathcal{D}
\ee
If the driving is noisy, then in general $G>\varrho_{\tbox{F}}(e/L)^2\mathcal{D}$. 
This reflects the simple observation that an insulator (${\mathcal{D}\sim 0}$)
may have a finite AC conductance (finite $\dot{\mathcal{W}}$).

\sect{(3)}{The quantum LRT calculation:}
Eq.(\ref{e4}) is formally valid also in the quantum mechanical 
case, and can be regarded as the outcome of 
the Fermi-Golden-rule picture. However one 
should use the proper expression for $\tilde{C}(\omega)$, 
taking into account the levels statistics 
and the level broadening:
\be{7}
\tilde{C}_{\tbox{qm}}(\omega) \ \ &=& \ \  R(\omega)\tilde{C}_{\tbox{cl}}(\omega)
\\ \label{e8}
\tilde{C}(\omega) \ \ &\approx& \ \  \delta_{\Gamma}(\omega){\star}\tilde{C}_{\tbox{qm}}(\omega)
\ee
where $R(\omega)$ takes care 
for the level statistics, 
$\delta_{\Gamma}(\omega)$
is the line shape of the level broadening,  
and $\star$ indicates a convolution.
It should be clear that with pure DC driving Eq.(\ref{e4}) 
gives zero due to the discreteness of the energy spectrum,  
unless there is some mechanism that ``broadens the levels" 
so as to have effectively a continuum. 
It is customary to express the quantum version 
of the Kubo formula using the matrix elements of the velocity operator:
\be{9}
G = \pi\hbar 
\left(\frac{e}{L}\right)^2
\sum_{n \neq m}
|v_{mn}|^2  
\ \delta_{T}(E_n{-}E_F) 
\ \delta_{\Gamma}(E_m{-}E_n)
\ee
The $T$ broadened delta should be interpreted 
as the derivative of the Fermi occupation function, 
while the functional shape of the $\Gamma$ broadened 
delta function will be discussed later.

\sect{(4)}{The quantum SLRT calculation:}
A loose way to write the last expression is 
\be{10}
G = \pi\hbar 
\left(\frac{e}{L}\right)^2
\varrho_{\tbox{F}}^2 
\,\, \langle\langle|v_{nm}|^2\rangle\rangle
\ee
The interpretation of ${\langle\langle...\rangle\rangle}$ 
within the framework of the tradition LRT 
is implied by Eq.(\ref{e9}). 
It involves the specification 
of the level broadening parameter $\Gamma$  
and of the occupation temperature $T$.

The implicit assumption in LRT is that the 
driving induced transitions are much 
slower compared with the environmentally 
induced relaxation. In other words it is 
assumed that the relaxation is very effective 
in ``killing" any quantum effect that goes 
beyond first order perturbation theory.
SLRT, unlike LRT is aimed in taking account 
also the opposite circumstances in which  
the possibility to make connected sequences 
of transitions becomes essential in order 
to get absorption \cite{kbr}. 
Following \cite{slr,bls} we regard the energy levels 
as the nodes of a resistor network. We define 
\be{11}
\mathsf{g}_{nm}  = 
2\varrho_{\tbox{F}}^{-3} \ 
\frac{|v_{nm}|^2}{(E_n{-}E_m)^2} \   
\delta_{\Gamma}(E_m{-}E_n)
\ee
Then it is argued that $\langle\langle|v_{nm}|^2\rangle\rangle^{-1}$ 
is the resistivity of the network. 
It is a simple exercise to verify  
that if all the matrix elements are the same, 
say  $|v_{nm}|^2 = \sigma^2$, 
then $\langle\langle|v_{nm}|^2\rangle\rangle = \sigma^2$
too. But if the matrix is structured or sparse 
then $\langle\langle|v_{nm}|^2\rangle\rangle$ is 
in general much smaller compared with the RMS value 
of the matrix elements.

\sect{(5)}{Level broadening:}
Neither LRT nor SLRT are self contained without 
the specification of the level broadening. 
For the purpose of analysis we can regard 
the environmental fluctuations as a noisy 
driving source. This driving source induces  
decoherence: It can be argued (see for example Ref.\cite{qkr}) 
that the effect of decoherence is to  
multiply the velocity-velocity correlation 
function by an exponential factor $\exp(-\Gamma|t|)$. 
This means that 
\be{12}
\delta_{\Gamma}(\omega) 
\ \ \approx \ \  
\frac{1}{\pi} \frac{\Gamma}{\omega^2+\Gamma^2} 
\ee
It can be further argued \cite{qkr}
that an estimate of the rate~$\Gamma$ 
can be obtained using the Fermi golden rule (FGR):
\be{13}
\Gamma = 
\int_{-\infty}^{\infty}  
d\omega \,
\tilde{F}(\omega) \, 
\tilde{C}_{\tbox{qm}}(\omega)
\ee
where $\tilde{F}(\omega)$ is the power 
spectrum of the environmentally induced noise.
At low temperatures (see later) 
the noise power spectrum is effectively  
very narrow. Schematically we can write  
${\tilde{F}(\omega) = \epsilon^2 \delta_T(\omega)}$, 
where $\epsilon$ is the noise amplitude, 
and $T$ is the temperature.

The FGR expression  Eq.(\ref{e13})
parallels Eq.(\ref{e4}), 
and can be expressed using the matrix 
elements of the perturbation as in Eq.(\ref{e9}), 
or it can be written loosely in the style of Eq.(\ref{e10}): 
\be{0}
\Gamma = \epsilon^2 \times 2\pi\hbar 
\left(\frac{e}{L}\right)^2
\varrho_{\tbox{F}} 
\,\, \langle\langle|v_{nm}|^2\rangle\rangle
\ee
The latter version is common in elementary textbooks 
(the FGR calculated decay rate is proportional to 
the squared matrix element times the density of states). 
It is implicit in this expression that 
the FGR transitions ${\delta(E_m{-}E_n)}$  
have a width (broadening) which is determined 
by the normalized power spectrum 
of the environmental fluctuations. 

So far we simply re-stated the standard textbook 
version of FGR using our notations.  
But it should be clear that the naive application 
of the FGR recipe is as problematic as the the naive 
application of the Kubo formula \cite{decay}:  
the first order (transient) transitions 
to ``neighboring" states are not enough
in order to get a non-transient decay. 
Rather, the possibility to have {\em connected sequences}
of transitions is essential in order to 
have a long time decay.  Thus we conclude     
that an analogous SLRT recipe, 
with $\mathsf{g}_{nm}$ defined in 
a similar fashion as in Eq.(\ref{e11}),  
should be applied in order to determine 
the long time value of $\Gamma$.
We point out again that in the calculation 
of~$\Gamma$ the role of the broadened delta 
is played by the normalized power spectrum 
of the environmental fluctuations. 
This means that at low temperatures (see further 
discussion in a later paragraph) we have  
in the $\Gamma$~oriented version of Eq.(\ref{e11}) 
a thermally broadened delta function ${\delta_T(E_m{-}E_n)}$.

\sect{(6)}{Moderate disorder:}
The effect of hard chaos or disorder is to 
randomize the velocity. Within the framework 
of the semiclassical LRT calculation we always 
get the Drude formula irrespective of the strength 
of the disorder. Following Drude it is 
customary to write 
\be{14}
\langle v(t) v(0) \rangle \Big|_{\tbox{cl}} = 
v_{\tbox{F}}^2 \exp\left[-2\left(\frac{v_{\tbox{F}}}{\ell}\right)|t| \right]
\ee
or equivalently 
\be{15}
\tilde{C}_{\tbox{cl}}(\omega) = v_{\tbox{F}}^2 \,\, 
\frac{(2v_{\tbox{F}}/\ell)}{\omega^2 + (2v_{\tbox{F}}/\ell)^2}
\ee
This leads to the Drude result
\be{16}
G = \frac{e^2}{2\pi\hbar} \mathcal{M} \frac{\ell}{L}
\ee
Once we turn to the quantum calculation,  
we have to be more careful. For moderate 
disorder the eigenfunctions are ergodic, 
and therefore the distinction between 
the LRT recipe and the SLRT recipe is not important.   
Furthermore, using a random wave conjecture 
for the eigenstates, one recovers  
the Drude expression $\tilde{C}_{\tbox{cl}}(\omega)$
as an approximation for the quantum $\tilde{C}(\omega)$.
In order to do a better job, the spectral function $R(\omega)$ 
is introduced. This function takes into account 
the level statistics: for large~$\omega$ it equals unity,
while for small~$\omega$ it reflects 
the repulsion between levels.
The ``level broadening" effect is taken 
care of by the convolution with $\delta_{\Gamma}(\omega)$, 
as indicated in Eq.(\ref{e8}). 
The standard LRT regime is having 
$\Delta \ll \Gamma \ll \Delta_b$. 
In this regime the introduction of $R(\omega)$ 
implies so-called weak localization corrections 
to the Drude result.  
These are found to be of order $\Delta/\Gamma$. 
See Refs.\cite{G1,G2,kamenev}, 
and also \cite{ophir} 
for the ``quantum chaos" point of view.

\sect{(7)}{Weak disorder:}
The SLRT recipe for the calculation 
of $G$ can be regarded as an extension 
of the LRT recipe. The results of SLRT 
become very different from those 
of LRT once the perturbation matrix $v_{nm}$ 
is either structured or sparse. 
The case of weak disorder (ballistic case) 
has been analyzed in a previous Letter \cite{bls}.
In the non trivial ballistic regime
($1 \ll \ell/L \ll \mathcal{M}$)  
each eigenfunctions occupy a large but 
finite fraction of open modes.
One may say that the eigenfunctions 
are ``localized" in mode space.
This lack of quantum-ergodicity 
implies sturctures and sparsity.  
Consequently the SLRT result is not 
merely a small weak localization correction:   
the leading order result is no longer Drude. 
In what follows we address the other 
extreme case of strong disorder 
(Anderson localization case). Also here 
we are going to see that the leading 
order result is not Drude.

\sect{(8)}{Strong disorder:}
The first step is to figure out how $R(\omega)$ 
look like for a system with strong localization. 
The initial spread of a wavepacket is diffusive 
with ${\langle (x(t)-x(0))^2 \rangle = 2\mathcal{D}_0t}$
where ${\mathcal{D}_0= v_{\tbox{F}} \ell}$. 
But for long time $\langle (x(t)-x(0))^2 \rangle = \ell_{\xi}^2$, 
where $\ell_{\xi}$ is the localization length. 
This implies a breaktime at ${t^* = \ell_{\xi}^2 / \mathcal{D}_0 }$, 
and therefore the velocity-velocity correlation
function is modified in the frequency range $|\omega| < (1/t^*)$.
Namely, the above implies 
\be{17}
R(\omega) \Big|_{\tbox{global}}
\ \ \approx \ \ 
\frac{1}{1+(t^*\omega)^{-2}}
\ee
Further considerations which are based on Mott's 
picture of resonances imply an extra  
$|\log\omega|^{d{+}1}$ factor at small frequencies.
We can argue in advance that this type of correction has 
no importance in SLRT because it reflects 
a very sparse contribution to the 
perturbation matrix, that cannot lead 
to connected sequences of transitions.

Even if we eliminate Mott's resonances, still 
the low frequency behavior of $\tilde{C}(\omega)$
reflects a very sparse matrix $|v_{nm}|^2$.
Within the framework of SLRT the sparse component 
of $|v_{nm}|^2$ does not contribute to the diffusion. 
Only the non-sparse component allows 
{\em connected sequences of transitions}.
We argue that the non sparse component is obtained 
by multiplying $\tilde{C}_{\tbox{cl}}(\omega)$ by  
\be{18}
R(\omega) \Big|_{\tbox{effective}}
\ \ \approx \ \
\exp\left[
-\left(
\frac{\Delta_{\xi}}{|\omega|}
\right)^{1/d} 
\right] 
\ee
The reasoning is as follows:
The matrix elements of states~$n$ that
are located within range $|x_n-x_m|<r$
from a given state~$m$ constitute 
a connected grid with spacing 
${\Delta_r = (L/r)^d\Delta}$
and typical value 
${|v_{nm}| \propto \exp(-r/\ell_{\xi})}$.
From the equation ${\Delta_r = \omega}$
we deduce that the "volume" of states
that contribute a connected grid for
$\omega$~transitions has a radius 
${r = (\Delta/\omega)^{1/d}L}$.
Hence we deduce the above formula,
where ${\Delta_{\xi} = (L/\ell_{\xi})^d\Delta }$.

\sect{(9)}{The hopping picture:}
Assuming that the coherence 
time $\tau_{\gamma}=1/\Gamma$ 
is much smaller compared with 
the breaktime~$t^*$ we observe 
that the $d\omega$ integral 
of Eq.(\ref{e4}) with (\ref{e7}-\ref{e8}) 
and the Lorentzian of (\ref{e12}) 
is dominated by the tail [$\omega>(1/t*)$] 
leading to the result
\be{19}
\mathcal{D} \ \ \approx \ \ 
\Gamma t^* \, \mathcal{D}_0
\ \ = \ \ \frac{(\ell_{\xi})^2}{\tau_{\gamma}}  
\ee
This is as expected from heuristic considerations. 
It describes a random walk hopping process with 
steps of size $\ell_{\xi}$ and time $\tau_{\gamma}$.  

The issue is to obtain an explicit expression 
for~$\Gamma$. Thermal noise is ``white" at 
high temperatures (${F(\omega) \propto T}$), 
with a very large temperature-independent cutoff frequency. 
In such case Eq.(\ref{e13}) implies $\Gamma \propto T$ 
which is not very interesting.  
Low temperature noise, unlike white noise, 
has an exponentially decaying emission tail,
and consequently, it has effectively 
a very narrow span of frequencies: 
\be{20}
F(\omega)
\ \ \sim \ \   
\exp\left(-\frac{|\omega|}{T}\right)
\ee
In the above expression we have 
neglected $\omega^{\alpha}$ term 
whose exponent depends of the detailed 
spectral properties of the bath.  
The calculation of $\Gamma$ with Eq.(\ref{e13})  
involves a $d\omega$ integral over
\be{21}
\exp\Big[
-\frac{|\omega|}{T}
\Big] 
\times 
\exp\Big[
-\left(
\frac{\Delta_{\xi}}{|\omega|}
\right)^{1/d} 
\Big] 
\ee
This is mathematically equivalent to the VRH integral \cite{mott2}.
There the optimization is over the range 
of the hopping~$r$, while here it is over 
the associated frequency~$\omega$. 
The optimal frequency is $\omega\propto T^{d/(d+1)}$ 
leading to the VRH estimate 
$\mathcal{D} \propto \exp[(-T_0/T)^{1/(1+d)}]$ 
where $T_0$ is a constant. It should be clear 
that the VRH estimate is an approximation for 
the resistor network calculation with Eq.(\ref{e11}). 
The VRH estimate works quite well for $d>1$, 
but gives the wrong exponent ($T^{-1/2}$) for $d=1$.
The correct exponent ($T^{-1}$) in the latter case 
is implied by the absence of percolation.

\sect{}{Conclusions:}  
We have established that SLRT 
provides a firm unified framework for the calculation 
of the conductance. Ad hock phenomenology is not  
required in order to establish the  
resistor network ``hopping" picture, 
from which Mott's VRH approximation is derived.
It should be clear that the generalized resistor network 
picture of SLRT is not limited to the strong disorder  
regime: it allows {\em on equal footing} the calculation  
of the conductance in the other extreme case of ballistic motion.   
The importance of this approach is also in its 
potential capabilities:  being a natural 
extension of LRT it also allows, in principle, 
the incorporation of many-body effects. In the latter case 
the calculation of conductance is reduced, as in LRT, 
to the analysis of the matrix elements of the current operator, 
whatever are the interactions involved.  

The long standing puzzle regarding ``AC conductance" 
versus ``DC conductance" that has been discussed in the 
introduction is automatically resolved by our theoretical framework. 
We claim that there is no crossover from ``AC conductance" 
to ``DC conductance" as a function of frequency. 
Rather, we argue that there is a crossover from ``AC conductance" 
to ``DC conductance" as a function of the driving intensity.
Using the terminology of Ref.\cite{kbr}, 
the crossover is from the ``spectroscopic" to the ``mesoscopic" 
result for the conductance: The Kubo formula of LRT 
applies to a spectroscopic measurement of the conductance, 
where the driving is assumed to be very weak compared 
with the relaxation processes; The hopping picture that emerges 
from SLRT applies in the mesoscopic regime where the relaxation 
is assumed to be slow compared with the driving-induced transitions.

\sect{}{Acknowledgment:}
The apparent inconsistency between the hopping/VRH picture 
and the Kubo AC/DC formalism, and the question how to reconcile 
between them, have been pointed out by Michael Wilkinson 
in the late nineties. 
The issue has been raised again in his 2005 visit in BGU, 
while discussing the extension of the Kubo formalism that 
has been presented in Ref.\cite{kbr}. This discussion 
has initiated further refinement of the theory \cite{slr}
in collaboration with him and with Bernhard Mehlig.
The belief that the theory can give hopping/VRH
has been shared by all of us and was implicitly expressed   
in \cite{slr}. Prior to submission of this manuscript I
have been notified that during the last year MW and BM 
have made independently further progress on SLRT in general 
and possibly on this issue in particular (undocumented).
The research was supported by a grant from the DIP, 
the Deutsch-Israelische Projektkooperation.


\end{document}